\documentclass[12pt,preprint,natbib]{aastex}
\input epsf

\def\s2n{S^{\prime}/N}
\def\rh{$R_{\rm D}$--$R_{\rm H}$}
\def\av{$\langle \tau_{\rm v} \rangle$}
\def\aav{$A_{\rm V}$}
\def\obsav{$A_{\rm V,obs}$}
\def\oav{$A_{\rm V,0}$}
\def\xe{$x_{\rm e}$}
\def\ratioy{[HCO$^+$]/[CO]}
\def\ratiox{[DCO$^+$]/[HCO$^+$]}
\bibpunct{(}{)}{;}{a}{}{,}

\begin{document}
\title{Electron Abundance in Protostellar Cores}

\author{Paolo Padoan\footnote{Currently at Physics Department, University of California, San Diego, La Jolla, CA 92093}, Karen Willacy, William Langer,} 
\affil{Jet Propulsion Laboratory, 4800 Oak Grove Drive, MS 169-506,
     California Institute of Technology, Pasadena, CA 91109-8099, USA}
\author{Mika Juvela} 
\affil{Helsinki University Observatory, T\"ahtitorninm\"aki, P.O.Box 14,
SF-00014 University of Helsinki, Finland}

\begin{abstract}

The determination of the fractional electron abundance, \xe, in protostellar
cores relies on observations of molecules, such as DCO$^+$, H$^{13}$CO$^+$
and CO, and on chemical models to interpret their
abundance. Studies of protostellar cores have revealed significant
variations of \xe\ from core to core within a range
$10^{-8}<x_{\rm e}<10^{-6}$. The physical origin of these large
variations in \xe\ is not well understood, unless unlikely variations
in the cosmic ray ionization rate or ad hoc values of metal depletion
are assumed. In this work we explore other potential causes of these 
variations in \xe\, including core age, extinction and density. We 
compute numerically the intensity of the radiation field 
within a density distribution generated by supersonic turbulence. Taking into 
account the lines of sight in all directions, the effective visual extinction 
in dense regions is found to be always much lower than 
the extinction derived from the column density along a fixed line of sight. 
Dense cores with volume and column densities comparable to observed 
protostellar cores have relatively low mass--averaged visual extinction, 
2~mag~$\le A_{\rm V}\le 5$~mag, such that photo--ionization can sometimes 
be as important as cosmic ray ionization.
Chemical models, including gas--grain chemistry and time dependent gas 
depletion and desorption, are computed for values of visual extinction 
in the range 2~mag~$\le A_{\rm V}\le 6$~mag, and for a hydrogen gas density 
of 10$^4$~cm$^{-3}$, typical of protostellar cores. The models presented
here can reproduce some of the observed variations of ion abundance from core to core 
as the combined effect of visual extinction and age variations.  The range
of electron abundances predicted by the models is relatively insenstive
to density over 10$^4$ to 10$^6$ cm$^{-3}$.

\end{abstract}


\section{Introduction}

The fractional electron abundance, 
$x_{\rm e}=n_{\rm e}/n_{\rm H_2}$, is an important physical parameter in the
chemistry of molecular gas because it controls the molecular ion abundance of 
the gas. The molecular ion abundance determines the coupling between 
the gas and the magnetic field and is therefore important for 
magneto--hydrodynamic (MHD) processes in molecular clouds. 
The ambipolar drift velocity of electrically 
neutral gas relative to charged particles (and magnetic field lines)
can be expressed as $v_{\rm AD}\propto B^2/x_{\rm e}$, where $B$ is the
magnetic field strength (assuming equilibrium of Lorentz
and frictional forces).

The determination of \xe\ in protostellar cores relies on the
observations of molecular ions, such as DCO$^+$ and 
H$^{13}$CO$^+$ and on chemical models to interpret the
abundance of such ions in relation to the value of \xe .
This method to determine \xe\ has been applied
by Caselli et al.\ (1998) and Williams et al.\ (1998), using 
updated deuterium abundance and dissociative recombination
rates of H$_3^+$ and H$_2$D$^+$, and complementary determinations
of column density through observations of the transition J=1--0 of
C$^{18}$O. Observational results and predictions of the chemical
models are compared on the \rh\ plane, where $R_{\rm H}=$\ratioy\ 
and $R_{\rm D}=$\ratiox\ are molecular abundance ratios.        

\nocite{Caselli+98} \nocite{Williams+98}

Significant variations from core to core in the value of 
$R_{\rm H}$ are found ($10^{-5}<R_{\rm H}<10^{-4}$), implying 
variations of \xe\ in the range $10^{-7.5}<x_{\rm e}<10^{-6.5}$,
according to Williams et al.\ (1998), and $10^{-8}<x_{\rm e}<10^{-6}$ 
in the study by Caselli et al.\ (1998). The physical origin of these 
large variations of \xe\ is not well understood. 
The observed HCO$^+$ abundances are matched by Caselli et 
al.\ (1998) by varying the cosmic ray ionization rate in the range 
$10^{-18}$~s$^{-1}\le \zeta_{\rm H_2}\le 10^{-16}$~s$^{-1}$, 
and by Williams et al.\ (1998) by varying the metal depletion 
by a factor of 2--15 times their nominal values.
However, the cosmic ray ionization rate has been 
measured to be $\zeta_{\rm H_2}=2.6\pm1.8\times 10^{-17}$~s$^{-1}$
within massive protostellar envelopes, where photo--ionization
is certainly insignificant (van der Tak \& van Dishoeck 2000). 
Variations of two orders of magnitude in $\zeta_{\rm H_2}$ are not
expected inside dense cores. Furthermore, there is no observational 
evidence for the specific values of metal depletion inferred by Williams
et al.\ (1998); as those authors point out, gas--grain models 
(Bergin, Langer \& Goldsmith 1995) predict a rather rapid metal
depletion, and metal abundances significantly lower than assumed 
in their work are expected, decreasing as a function of age.
This result is also confirmed by the models presented in this work
(see \S~5).

\nocite{vanderTak+vanDishoeck2000} \nocite{Bergin+95}

Here we investigate the possibility that the observed 
abundances of molecular ions and the inferred variations of 
\xe\ in protostellar cores are due to variations in age and 
extinction, $A_{\rm V}$, from core to core.
In \S~2 we show, based on numerical simulations, that dense 
cores formed by supersonic turbulent flows within molecular clouds can 
be much less opaque to UV photons than inferred from their observed column 
density, due to the very complex density distribution in the cloud and 
within the cores. In this work $\tau_{\rm v}$ is the effective visual 
extinction in three dimensional models, computed with the Monte Carlo 
radiative transfer code. $A_{\rm V}$ is the visual extinction through 
the line of sight, proportional to the column density in that direction. 
$A_{\rm V}$ is also used to refer to the visual extinction in the chemical 
models. Values of effective visual extinction, $\tau_{\rm v}$ , 
are found to be smaller than the extinction measured in the 
direction of a fixed line of sight. 

In \S~3 we present chemical models, including gas--grain chemistry and 
time dependent depletion and desorption. The models are computed for a 
range of core visual extinction including the values suggested by the 
radiative transfer models, 
2~mag~$\le A_{\rm V}\le 6$~mag, and for values of gas density typical of 
protostellar cores (and consistent with the density in the cores selected 
in the turbulence simulations), $n= 10^4$~cm$^{-3}$. Models with
$n= 10^5$~cm$^{-3}$ and $n= 10^6$~cm$^{-3}$ are also computed. 
We show in \S~4 that the observed variations in ion abundances from core 
to core can be understood as the combined effect of variations in visual 
extinction and age (possibly gas density as well). Results are discussed 
in \S~5, and conclusions are summarized in \S~6.

\section{UV Flux in Turbulent Clouds}

The enhanced penetration of UV photons in clumpy or fractal 
molecular clouds has been discussed in previous works
(Boiss\'{e} 1990; Myers \& Khersonsky 1995; Elmegreen 1997). Here we compute 
the radiation field within a density distribution that results 
from the numerical solution of the magneto--hydrodynamic (MHD) equations, 
in a regime of highly supersonic turbulence. The model we use is the
final snapshot of the super--Alfv\'{e}nic simulation presented
in Padoan \& Nordlund (1999). We refer the reader to that 
paper for a detailed description of the simulation. The 
numerical experiment is run on a 128$^3$
staggered grid, with periodic boundary conditions, random
external large scale forcing and an isothermal equation of state.
Supersonic and super--Alfv\'{e}nic turbulent flows have been 
shown to provide a good description of the dynamics of 
molecular clouds (e.g. Padoan \& Nordlund 1999; Padoan et al.\ 
1999, 2001).

\nocite{Boisse90}
\nocite{Padoan+Nordlund99MHD}  
\nocite{Elmegreen97a} \nocite{Myers+Khersonsky95}
\nocite{Padoan+99per} \nocite{Padoan+2001cores} 

The intensity of the radiation field is computed with a Monte Carlo 
method. Photon packages are sent into the cloud from the background and
the scattering and absorption processes are simulated. The dust opacity is 
calculated from the relation between $N_{\rm H_2}$ and $A_{\rm V}$ as given 
by Bohlin, Savage \& Drake (1978). The albedo of the grains is assumed to be 
0.5 and the asymmetry factor of the scattering $g=0.6$ (e.g. Draine \& Lee 1984).

\nocite{Draine+Lee84}

The number of incoming photons is obtained for each cell at position ${\bf r}$, 
and the local value of the visual extinction, $\tau_{\rm v}({\bf r})$, is computed
from the local intensity of the radiation field, $I({\bf r})$, relative to 
external radiation field, $I_0$:
\begin{equation}
\tau_{\rm v}({\bf r}) = 2.5\,{\rm log}_{10}({\rm e})\,\tau({\bf r})
\label{ex1}
\end{equation}
where $\tau({\bf r})$ is the effective optical depth at the position ${\bf r}$,
\begin{equation}
\tau({\bf r})=-{\rm ln}\left[ \frac{I({\bf r})}{I_0}\right]
\label{ex2}
\end{equation}
and 
\begin{equation}
I({\bf r})=\frac{1}{4\pi}\int{I({\bf r},\omega)\,d\omega}
\label{ex3}
\end{equation}
where $I({\bf r},\omega)$ is the radiation field at the position ${\bf r}$ 
from the direction $\omega$. 

\nocite{Bohlin+78}

Figure~1 shows volume projections of the density field
(left panel) and visual extinction (right panel) in our 
model. Dense cores and filaments can be seen in the 
density field. The visual extinction increases
gradually toward the center of the model and also inside
the densest cores and filaments. Therefore, both the local
density and the large scale structure affect the local value 
of the extinction. 

The local value of $\tau_{\rm v}$ in the three dimensional model 
cannot be derived directly from the visual extinction through 
the line of sight, $A_{\rm V}$, corresponding to the column density 
along that line of sight. In Figure~2, left panel, the largest value 
of the local visual extinction, $\tau_{\rm v,max}$, 
along each of the lines of sight parallel to the three orthogonal axis 
of the numerical mesh, is plotted versus the 
visual extinction derived from the column density along 
that line of sight divided by two, \obsav . The values of 
$\tau_{\rm v,max}$ are always much smaller than the values
of \obsav . The right panel of Figure~2 shows a linear--log plot
of the histograms of the two quantities. 

The result illustrated in Figure~2 implies that column density 
determinations cannot provide useful estimates of the actual 
$\tau_{\rm v}$ in dense regions of turbulent molecular clouds. 
More importantly, local values of $\tau_{\rm v}$ are likely to be a few 
times smaller than the observed \obsav . This difference is due to the 
existence of directions with optical depth smaller than the specific line 
of sight of the observations and to the presence in the cloud of diffuse 
scattered radiation. This result should be taken into account when 
dust models are used to constrain dust temperature and emissivity in 
cores and when chemical models are used in combination with extinction 
measurements and chemical abundance determinations to constrain
gas depletion.

We select dense cores in the MHD density snapshot as 
connected regions with gas density above 10$^4$~cm$^{-3}$.
This value is chosen to match the typical densities of
protostellar cores and also for consistency with the 
observational sample in Williams et al.\ (1998), where 
cores are selected from their ammonia emission and have a
typical density of approximately 10$^4$~cm$^{-3}$.
We compute the mass--averaged visual extinction in each core, 
\av , assuming it is the appropriate value of visual 
extinction that should be used in single point chemical models 
to derive molecular abundances in the cores (the observed
abundances are also approximately mass--averaged, given the 
low opacity of the observed transitions of DCO$^+$, 
H$^{13}$CO$^+$ and C$^{18}$O). The mass--averaged visual 
extinction also has the advantage of being only weakly 
dependent on the density threshold used to select the cores.

We compute the total mass and volume of each core, and use them 
to derive the core column density assuming a spherical shape, as an 
estimate of the average core column density. The visual extinction to
the center of each core corresponding to this value of column density
is called \oav . As illustrated in Figure~3 (left panel), values of \av\ 
in dense cores are always significantly lower than the extinction derived 
from half the value of their column density, \oav .
The difference between values of \av\ in cores and the extinction 
derived from half the value of the column density at the line of sight
through the core center, \obsav , is even greater, as shown in the
right panel of Figure~3. This can be explained as due to the elongated
shape of cores in numerical simulations, consistent with typical
aspect ratios in observed cores (Myers et al.\ 1991), and to
density structures along the line of sight to a core and within
the core (lines of sight perpendicular to the elongated
structure have more weight in determining the value of $\tau_{\rm V}$).

Observed column densities almost always overestimate the 
core's true opacity to UV photons (especially at the largest values of
\obsav ), since these always penetrate preferentially along directions 
of lowest column density (for example the core shortest axis). The local 
extinction, $\tau_{\rm v}({\bf r})$, is significantly 
smaller than the extinction corresponding to the average optical depth, 
$\langle \tau({\bf r},\omega) \rangle_{\omega}$. This result can be easily 
understood in the absence of scattering or if the scattering is only forward. 
The local radiation field from the direction $\omega$ can then be expressed as 
a function of the local optical depth from that direction, 
$I({\bf r},\omega)=I_0\,{\rm exp}[-\tau({\bf r},\omega)]$, and the local
extinction is given by an average of ${\rm exp}[-\tau({\bf r},\omega)]$
The directions of lowest optical depth have therefore a dominant contribution 
to the average extinction.

\nocite{Myers+91}

Squares in Figure~3 represent cores at the distance from 
the cloud center smaller than $\langle R\rangle$, while asterisks represent
cores at a distance from the cloud center larger than $\langle R\rangle$,
where $\langle R\rangle$ is the average of the distances of all
selected cores to the cloud center (the center of the computational 
mesh). The figure shows that the visual extinction increases towards
the cloud center. Furthermore, the left panel shows that the extinction 
also increases with the cores mass. This shows again the visual 
extinction is determined both by the local mass distribution in an
individual core and by the cloud structure on a larger scale.

The cores we have selected from the MHD simulation have 
volume and column densities typical of observed protostellar 
cores. In the core sample by Williams et al.\ (1998), the column 
densities estimated by the intensity of the J=1--0 line
of C$^{18}$O correspond to \obsav\ in the range
$\sim$5--20~mag (the total extinction along the line of sight
divided by two), as in our numerical sample (see Figure~3, right 
panel). Since in our numerical sample we find corresponding values 
of \av\ in the range $\sim$2--5~mag (Figure~3), we 
can infer that the dense cores in the sample by Williams 
et al.\ (1998) may also have similarly low values of 
mass--averaged visual extinction, \av . 

The photoionization rate is larger than the cosmic ray ionization 
rate at $A_{\rm V}<4$~mag (McKee 1989). Equilibrium abundances of 
molecules and fractional ionization are sensitive to the value of 
$A_{\rm V}$ also for $A_{\rm V}\le 4$~mag. Based on the low values 
of \av\ predicted in this work, photoprocesses in protostellar cores 
could therefore be more important than assumed in previous studies.

\nocite{McKee89}

\section{Chemical Models of Cloud Cores}

The chemical model is based on \cite{Willacy+Millar98} and includes deuterium
reactions.  Gas phase molecules
freezeout onto dust grains at a rate of 
\begin{equation}
\frac{dn_{\rm x}}{dt} = C s_{\rm x} \langle \pi \, a^2 n_{\rm g} \rangle \,
v_{\rm x} n_{\rm x} \,\,\, {\rm cm}^{-3} {\rm s}^{-1}
\label{}
\end{equation}
where $s_{\rm x}$ is the sticking coefficient ($s_{\rm x}=0.3$ for all 
species), 
$a$ is the grain radius, $n_{\rm g}$ is the number density of grains (equal 
to $10^{-12} n_H$), $v_{\rm x}$ is the thermal velocity, and $n_{\rm x}$ 
is the number density of the species x. $C$ is a factor that accounts for
the expected increase of the freezeout rate of positive ions onto negative 
grains.  For neutrals $C=1$, while for ions $C=1+16.71\times 10^{-4}/(a \, T)$ 
(Rawlings et al.\ 1992).

\nocite{Rawlings+92}

We assume that H$^+$, H$_2^+$, H$_3^+$, their deuterated equivalents
and He$^+$ are neutralized on collision with grains and returned to the gas.
Desorption due to cosmic ray induced heating of grains is also 
included, using the rates given in Hasegawa \& Herbst (1993). The time evolution 
of the abundance of some molecular species is plotted in Figure~4. 
The elemental abundances are given in Table~\ref{t1} relative to 
$n(\rm H_{tot})$, where $n(\rm H_{tot})=2n({\rm H}_2)+n({\rm H})$.

\nocite{Hasegawa+Herbst93}

The main parameters of chemical models are the visual
extinction, $A_{\rm V}$, the gas temperature, $T$, the
gas density $n$, the cosmic ray ionization rate
for molecular hydrogen, $\zeta_{\rm H_2}$, the initial 
abundances of atomic species and the factor $\chi$
by which the interstellar radiation field is enhanced
relative to its standard value of 
$1.6\times 10^{-3}$~ergs~cm$^{-2}$~s$^{-1}$ (Habing 1968).   
In this work, we assume a temperature of $T=10$~K for both the 
gas and the dust, a cosmic ray ionization rate of 
$\zeta_{\rm H_2}=1.2\times10^{-17}$~s$^{-1}$ and a standard value
of the interstellar radiation field ($\chi=1.0$).

\nocite{Habing68}

Single point models are run for three values of molecular hydrogen gas density,
$n=10^4$, $10^5$ and $10^6$~cm$^{-3}$, and for nine values of 
extinction ($A_{\rm V}=2.0$, 2.5, 3.0, 3.5, 4.0, 4.5, 5.0, 5.5 and 6.0~mag)
at each density. 
As discussed above,
the values of visual extinction are taken from the MHD models, which 
are consistent with the column densities of the observational sample in 
Williams et al.\ (1998).
To simulate the evolution of gas under diffuse conditions
before it is incorporated into a core, we used the steady-state
output of a gas phase only model with density = 10$^{3}$ cm$^{-3}$ and
$A_v$ = 3 to provide the input abundances for the models presented
here.  The input elemental abundances for the gas phase model are
given in Table 1.

\section{Results}

In this section, results from the chemical models are compared with
observational data with the aim of determining whether variations
in the model parameters such as \aav, time and density can account
for the observed scatter in \ratiox\ and \ratioy\ . 
The observational \ratiox\ and \ratioy\ values 
are obtained by selecting only the starless cores from the samples 
in \cite{Williams+98} and Butner, Lada and Loren (1995). NH$_3$ and 
N$_2$H$^+$ abundances for a subsample of those starless cores are 
taken respectively from Jijina, Myers and Adams (1999) and from 
\cite{Caselli+2002cores}.
The densities of the cores in the sample of \cite{Williams+98}
lie in the approximate range of a few x 10$^3$ to 10$^{5}$ cm$^{-3}$.
Here we adopt a model density of 10$^{4}$ cm$^{-3}$ as our baseline
for comparison with the observations.  Changes in
model density cause some variation in the 
results, but by far the largest effects are caused by variations
in \aav\ or time.  

The time-dependent abundances of some
important species are shown in Figure~\ref{fig4} for a model
with $n = 10^4$ cm$^{-3}$ and \aav = 5.5 mag.
The abundance of CI drops off due to accretion and to gas phase conversion
into CO, but rises again slightly after a few $\times$ 10$^5$ years, where
the desorption of CH$_4$ due to cosmic ray heating provides a 
means by which a low level of carbon atoms can be maintained in the gas.
(The CH$_4$ is produced by hydrogenation of CI on grain surfaces).
As expected the abundance of HCO$^+$ and DCO$^+$ follows that of CO
and N$_2$H$^+$ begins to deplete at later times than CO. 

The main problem in estimating the electron abundance in protostellar
cores is the difficulty of observing directly some of the most
abundant molecular ions, such as H$_3^+$ and H$_3$O$^+$,
and atomic carbon and metals with low ionization potential.
Values of \xe\ are therefore determined
by matching the observed abundances of other molecular ions such as 
HCO$^+$ and DCO$^+$, with chemical models.  These molecular ions
are predicted to be the main carriers of charge at the
relevant core timescales.
The comparison between observed and predicted abundances is greatly
affected by the depletion of gas species on dust grains. Evidence
of CO depletion has been provided for many dense protostellar cores 
\citep{Willacy+98,Caselli+99,Tafalla+2002,Jessop+Ward-Thompson2001,
Bergin+2002,Hotzel+2002,Bacmann+2002,Caselli+2002,Juvela+2002depl}.
The depletion rate is directly proportional to the gas density
and models have shown that some molecules such as CO 
are removed from the gas more quickly than others e.g. N$_2$H$^+$,
NH$_3$ because of chemical effects (see Figure~\ref{fig4}).  As a result
cores are expected to show chemical differentiation with species
such as HCO$^+$ and DCO$^+$ being preferentially located in the
outer layers of cores where the density may be lower or the chemistry
is less evolved, and the denser, more evolved inner regions of the
cores being better traced by N$_2$H$^+$ and N$_2$D$^+$ (Kuiper et al.\ 1996,
Caselli et al.\ 2002b).  Spatial variations of abundances due to
depletion are a source of uncertainty in the comparison of single
point chemical models with the observations, even if the time
evolution of the depletion is modeled as in this work.  However
the combination of a variety of tracers can give clues as to
the chemical age of the cores.

Figure~\ref{fig10} shows the variation in abundance ratios of several molecules
with \aav\ for $n$ = 10$^4$ cm$^{-3}$.  Three model times are
shown: 4 $\times$ 10$^4$,
10$^5$ and 3 $\times$ 10$^5$ years.  We find that the abundances
at later times are not consistent with the observed values and
that the cores appear to be chemically young.
However for a fixed density and $A_v$ no single model can explain all
of the observations.  One possibility is that
the emission from different 
molecules may originate from different regions of the cloud 
as discussed above (c.f. Kuiper et al.\ 1996, Caselli et al.\ 2002b).  

\nocite{Jijina+99} \nocite{Butner+95} \nocite{Kuiper96}

Figure~\ref{fig5} shows the results for the model with n = 10$^4$ cm$^{-3}$
plotted on the \rh\ plane, where $R_{\rm D}$=\ratiox , and $R_{\rm H}$=\ratioy .
Squares show the \rh\ values of the observed starless cores.
The solid lines show the variations in abundance for different
values of \aav\ for fixed ages of 
$4\times 10^4$~yr, $10^5$~yr and
$3\times 10^5$~yr, with the bottom of the
curves corresponding to \aav = 3~mag and the
top to \aav = 6~mag. The filled diamonds correspond to \aav =3.5~mag. 
The free--fall time at this density is $t_{\rm ff}=3.5\times 10^5$~yr. 
Figure~\ref{fig5} shows that variations in age and extinction
can explain some, but not all, of the observed scatter in the \rh\ plane, at a fixed
density of $n=10^4$~cm$^{-3}$.  Increasing the density moves
the curves down slightly but not sufficiently to account for the
three cores in the bottom right hand corner of the \rh\ plane.

Figure~\ref{fig6} shows time evolution tracks of models with 
$n=10^4$~cm$^{-3}$ and \aav=3~mag, \aav=3.5~mag and \aav=6~mag
on the \rh\ plane. Each track is shown as a solid line for
an age up to $10^5$~yr, as a dashed line between $10^5$~yr and 
$2\times 10^5$~yr, and as a dotted line between $2\times 10^5$~yr 
and $10^6$~yr. Solid diamonds correspond to 
$t=t_{\rm ff}=3.5\times 10^5$~yr. As suggested by the isochrones,
some of the scatter in the observational \rh\ plane can be explained
by reasonable variations of age and visual extinction from core 
to core.  However, considering the observational error bars, five cores 
cannot be fit by our models: three of these have high \ratiox\ and low
\ratioy\ , one low \ratiox\ and high \ratioy\ and one both low \ratiox\
and low \ratioy\ . 
We find that rather young cores fit the observations, with ages 
smaller than $t_{\rm ff}$. 

Figure~\ref{fig6} also shows that, for fixed values of 
extinction, \aav $\ge 3.5$~mag, and density, $n=10^4$~cm$^{-3}$,
cores can span some of the  range of observed \ratiox\ and \ratioy\   
values for timescales of less than 0.2~Myr. Observed values of \ratiox\ 
have been recently used to estimate the ``chemical age'' of cloud cores in 
Taurus \citep{Saito+2002}. 

The same time evolution tracks from Figure~\ref{fig6} are plotted in 
Figure~\ref{fig7} on the planes \ratiox --\xe\ (top panel)
and \ratioy --\xe\ (bottom panel). The electron abundance
decreases with time, reaching a minimum at an age of 
approximately 0.2~Myr for the model with \aav =6~mag. As can be 
seen in Figure~\ref{fig8}, the electron abundance is anticorrelated with 
the abundance of molecular ions such as HCO$^+$, H$_3$O$^+$ and
H$_3^+$, as these ions are responsible for most recombinations. 
These molecular ions are also the main charge carriers for core ages 
greater than 0.1~Myr, while for younger cores the metals are the main charge 
carriers (Figure~\ref{fig8}). Metal ions are initially more abundant than
molecular ions because molecules are still being formed, and freezeout
has not yet removed the metals from the gas. 

Williams et al.\ (1998) estimated values of electron abundance
in the range $3\times 10^{-8}<x_{\rm e}<2\times 10^{-7}$ for the cores 
in their sample, excluding B133.  Our models give smaller values and a 
smaller range of electron abundances, 
$1.5 \times 10^{-8} <x_{\rm e}<7 \times 10^{-8}$.

The models of Williams et al.\ consider a gas phase only chemistry with
the effects of freezeout being modeled by considering variations in the
atomic carbon, oxygen and metal abundances.  
In contrast, our model includes explicit consideration
of the time-dependent depletion of the molecules from the gas.
The values of depletion we estimate for the observed cores are 
much larger than in Williams et al.\ (1998). The abundance
of Si$^+$ can be used for a comparison. Williams et al.\ (1998) find
solutions for most of the cores in their sample with Si$^+$ abundance
values in the range $2.7\times10^{-9}<\rm{[Si^+]/[H_2]}<2.0\times 10^{-8}$.
The highest depletion, $\rm{[Si^+]/[H_2]}=2.7\times10^{-9}$, corresponds
to the lowest electron abundance and the highest \ratioy\ value.
In the models used in this work the depletion of metals increases with time.
The initial Si$^+$ abundance is $\rm{[Si^+]/[H_2]}=4\times10^{-8}$.
At the free--fall time ($3.5\times 10^5$~yr for $n = 10^4$ cm$^{-3}$), 
$\rm{[Si^+]/[H_2]}=5.0\times10^{-10}$ 
in the model with \aav =6~mag and $\rm{[Si^+]/[H_2]}=4.4\times10^{-10}$ in
the model with \aav =3~mag. At an age of $t=1$~Myr the abundances for the same
two models are $\rm{[Si^+]/[H_2]}=3.4\times10^{-12}$ and 
$\rm{[Si^+]/[H_2]}=6.6\times10^{-14}$ respectively. Metal depletion is
therefore important even for core ages as low as one free--fall time.

According to the models used in this work, metal depletion is 10 to 10$^5$
times larger than assumed in Williams et al.\ (1998), even with cosmic
ray desorption. One of the reasons for the Williams et al.\ approach in
using only gas phase chemistry was that
they found that the vertical spread of their results in the \rh\ plane
disappeared with the inclusion of a gas-grain chemistry, as a result
of the rapid accretion of metal ions.  
In our models some vertical spread can be obtained due to both differences
in time and in \aav, but density has little effect.  
Although we are not able to reproduce as large a vertical spread as
Williams et al.\, the present 
models are still able to produce some of 
the observed scatter in \ratiox\ and \ratioy .

On the other hand, CO depletion is not very strong for ages up to 1~Myr,
as shown in Figure~\ref{fig4}. The CO abundance varies by less than a
factor of two between $t=0.1$~Myr and $t=0.5$~Myr. The uncertainty in the 
core C$^{18}$O column density values due to variations of the C$^{18}$O
abundance should be less than a factor of two, which does not  
affect significantly the results of this work.

Assuming that the electron abundance, \xe , is determined by 
cosmic ray ionization balanced by recombination, \xe\
can be expressed as
\begin{equation}
x_{\rm e}=C_{\rm i}\,n^{-1/2}\,\zeta_{\rm H_{\rm 2}}^{1/2}
\label{mckee}
\end{equation}
where $C_{\rm i}$ is a constant that depends on the relative contributions 
of molecular ions and metals to the ionization balance. 
This expression for \xe\ would be appropriate for cores where \aav\ $>$
4, where ionization due to cosmic rays dominates that resulting from UV
radiation.  McKee (1989) derives
$C_{\rm i}=3.2\times10^3$~cm$^{-3/2}$s$^{1/2}$ 
(dotted line in Figure~\ref{fig8}), while
Williams et al.\ (1998) obtained $C_{\rm i}=2.0\times10^3$~cm$^{-3/2}$s$^{1/2}$.
Both these values of $C_i$ produce \xe\ that is a factor of 3 -- 5 higher than 
predicted by the current models for $n = 10^4$ cm$^{-3}$
and \aav\ = 6 (Figure~\ref{fig8}).  Figure~\ref{fig8} also shows that
\xe\ varies with time by approximately a factor of 2.  Also if \aav\
is less than 4 mag, then the UV field will be important in determining
the value of \xe\, and since variations in \aav\ can account for some of the
spread in \rh\ a more complete consideration of the chemistry is required
to determine \xe .

\section{Conclusions}

This work presents a partial solution to the problem of the large scatter in the observed 
values of \ratiox\ and \ratioy\ in protostellar cores. It is shown that variations 
in \ratiox\ and \ratioy\ from core to core can be interpreted as the result of age 
and visual extinction variations. However, it is unable to explain
all of the variations, in particular those cores with high \ratiox\ and 
low \ratioy\ and the one core with low \ratiox\.
This interpretation also implies the following conclusions:
\begin{itemize}

\item The electron abundance in the observed protostellar cores is in the range
$1.5\times 10^{-8}<x_{\rm e}<7\times 10^{-8}$.

\item Protostellar cores in this sample are relatively young, with ages from a fraction of a 
free--fall time to one free--fall time.

\item A range of effective extinction between 2.5 and 5--6~mag is required by
the observations (the upper limit is not well defined since the chemical models are
not very sensitive to extinction for \aav $>5$~mag), consistent with the mass--averaged
extinction of dense cores in simulations of supersonic MHD turbulence.

\end{itemize}

This solution to the problem of the large scatter in the observed 
values of \ratiox\ and \ratioy\ in protostellar cores appears to be more
consistent with the observational data than the alternative solutions
proposed by Caselli et al.\ (1998) and Williams et al.\ (1998). Caselli 
et al.\ (1998) used a very large range of values for the cosmic ray ionization 
rate, which is not supported by observations of dense cores. Williams et al.\
(1998) used a very small value of metal depletion, which is not expected to
hold for dense cores, even in the presence of desorption mechanisms.
On the contrary, the present work is based on models with a realistic value 
of the cosmic ray ionization rate and a self--consistently determined
value of metal depletion.

\acknowledgements

We are grateful to Ted Bergin for his guidance in the preparation of this 
work and for his comments on a draft of this paper. Comments by the referee
also led to an improvement of this paper.
The work of PP was performed while he held a National Research
Council Associateship Award at the Jet Propulsion Laboratory,
California Institute of Technology. This research was supported in 
part by grants from NASA to JPL.

\clearpage


\clearpage

\onecolumn

{\bf Figure captions:} \\

\begin{figure}
\centerline{
\epsfxsize=11cm \epsfbox{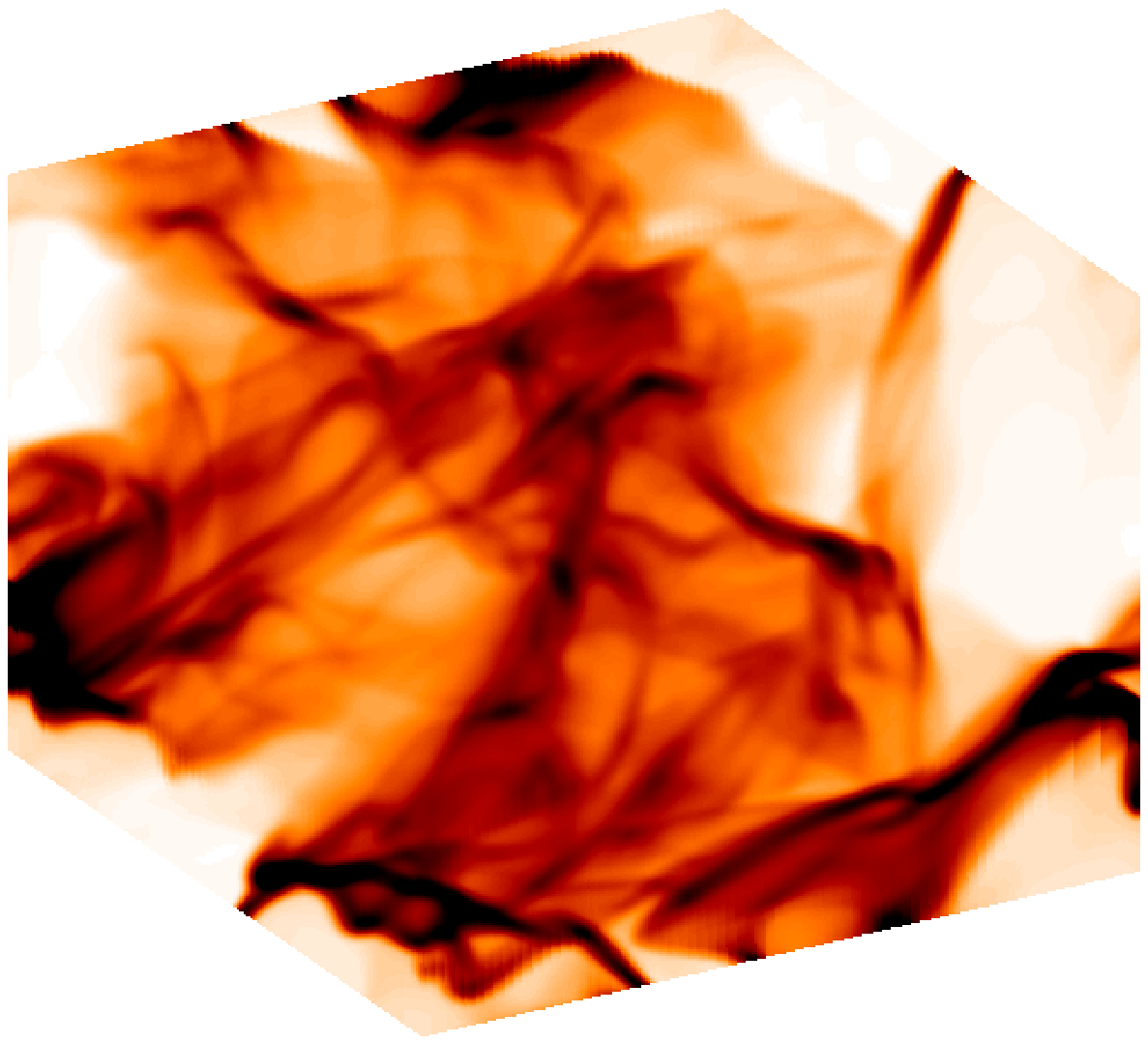}
\epsfxsize=11cm \epsfbox{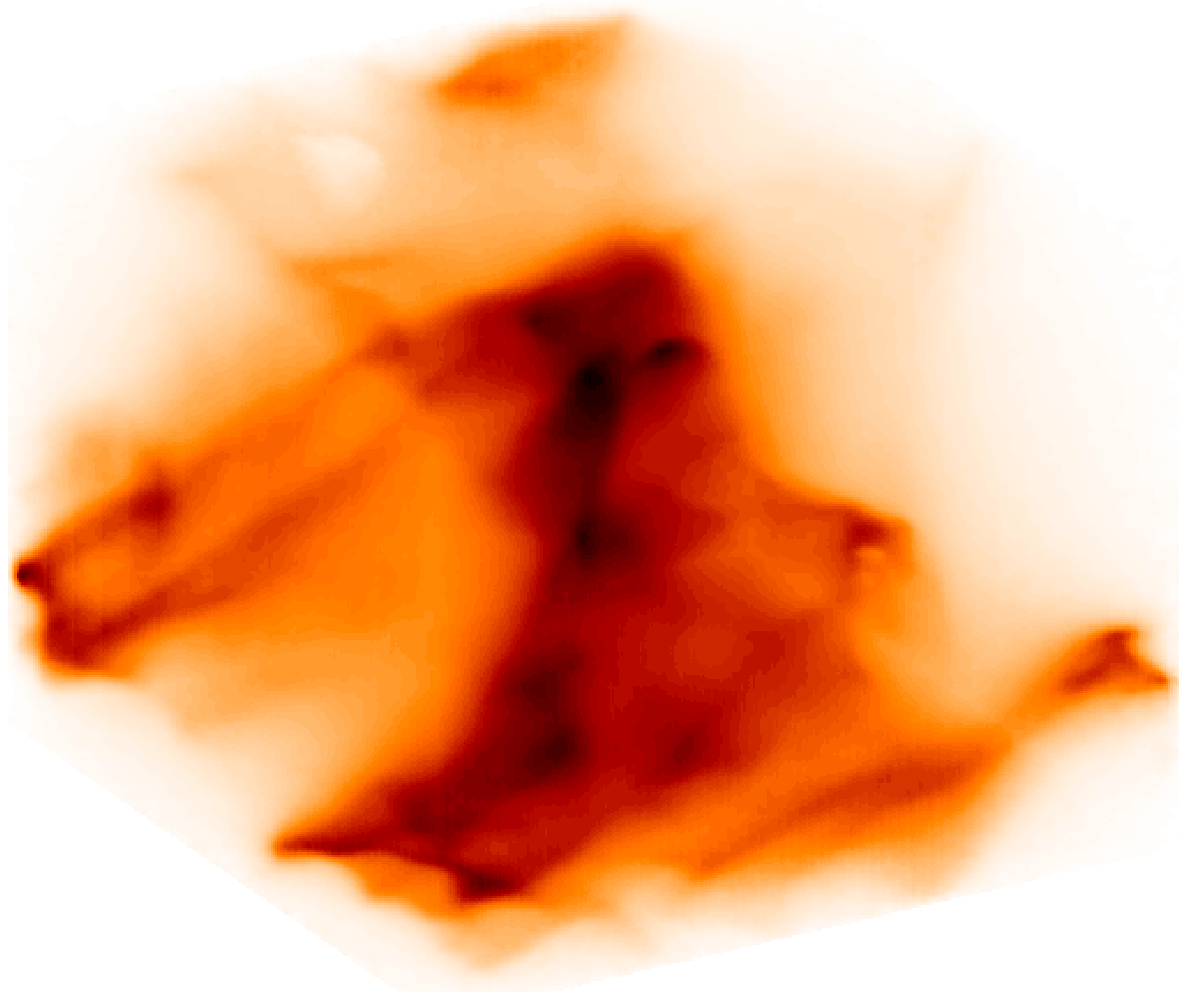}
}
\caption[]{}
\label{fig1}
\end{figure}

{\bf Figure \ref{fig1}:} Left panel: Volume projection of the density
field from the MHD simulation used in this work. Dark is large
density. Right panel: Volume projection of the effective visual extinction,
$\tau_{\rm v}$, assuming an isotropic radiation field outside of the computational mesh,
within the same mass distribution shown in the left panel. 
Dark is large visual extinction. \\

\begin{figure}
\centerline{\epsfxsize=19cm \epsfbox{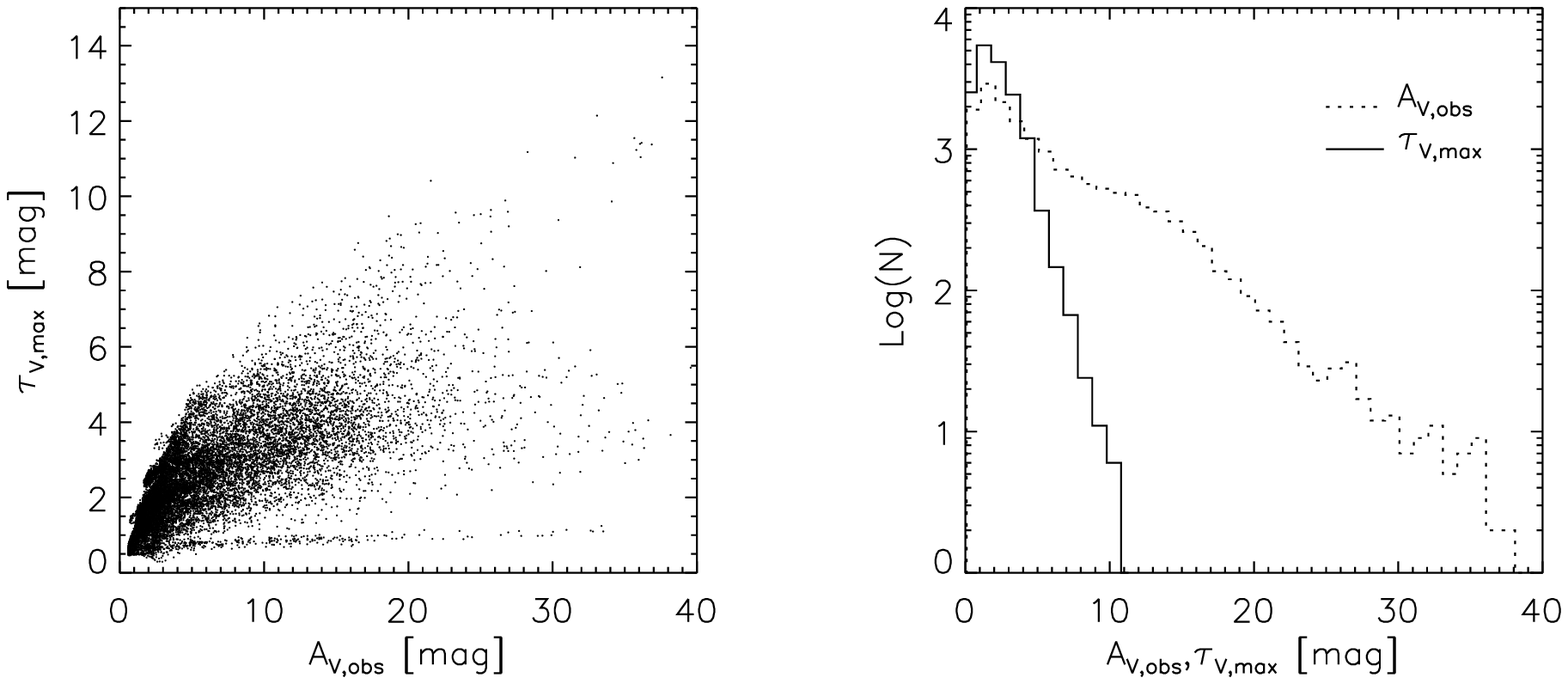}}
\caption[]{}
\label{fig2}
\end{figure}

{\bf Figure \ref{fig2}:} Left panel: Scatter plot of the maximum 
extinction within a line of sight, $\tau_{\rm v,max}$, versus the 
extinction due to half of the total column density of that line of
sight, \obsav . Right panel: Lin--log plot of the histograms of the
two quantities plotted in the left panel. \\

\begin{figure}
\centerline{\epsfxsize=17cm \epsfbox{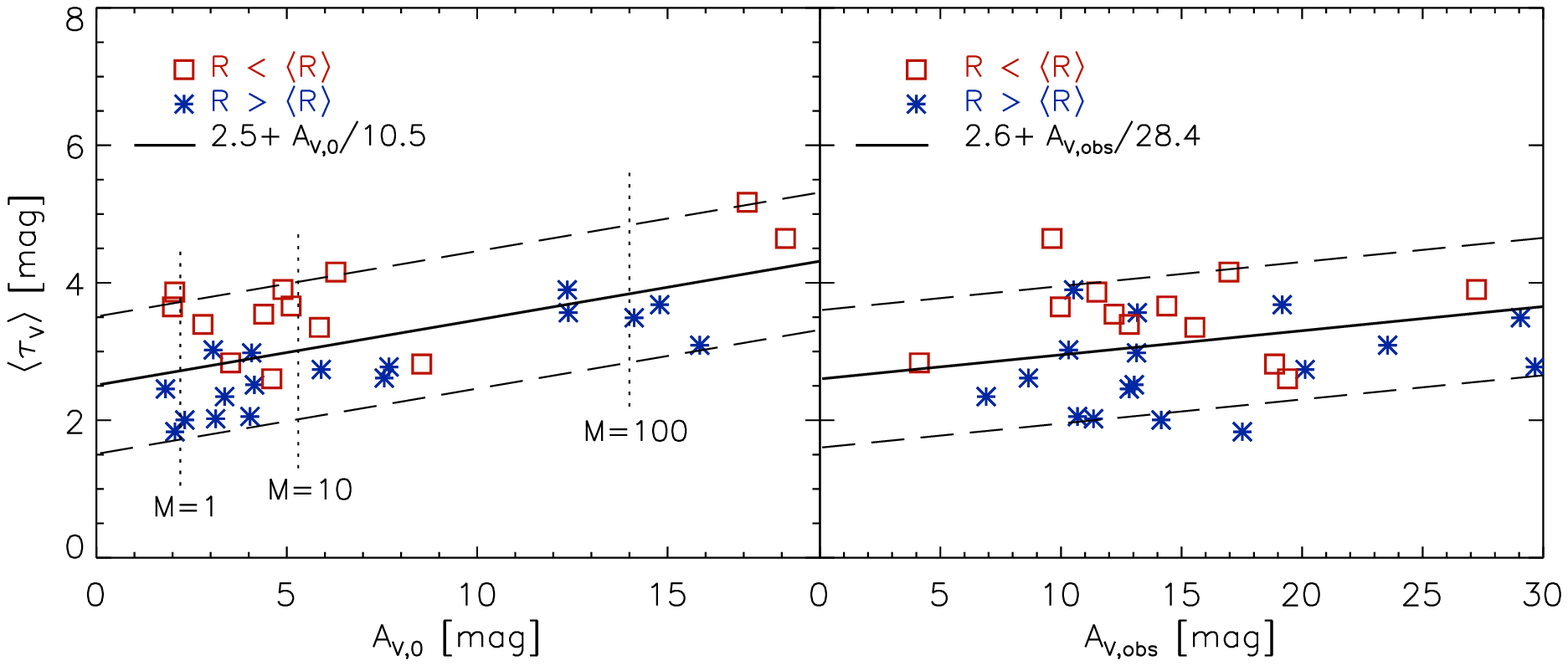}}
\caption[]{}
\label{fig3}
\end{figure}

{\bf Figure \ref{fig3}:} Left panel: Mass averaged visual extinction,
\av, within the model cores selected from the MHD simulation, versus
their extinction derived from their mass assuming spherical uniform density
distribution, $A_{\rm V,0}$. The respective core mass values range from 
approximately 1 to 200 solar masses. $\langle R \rangle$ is the average
distance of cores to the cloud center. Cores near the cloud center
($R<\langle R \rangle$) have larger visual extinction than cores
closer to the cloud surface ($R>\langle R \rangle$).
Right panel: Mass--averaged visual extinction,
\av, within the model cores selected from the MHD simulation, versus
the extinction due to half of the total column density along the line of
sight through the same cores, \obsav . The model cores have values of 
\obsav\ comparable to the ones found in the protostellar core sample by
Williams et al. (1998), but their mass--averaged effective extinction, \av, is
significantly smaller.\\

\begin{figure}
\centerline{\epsfxsize=20cm \epsfbox{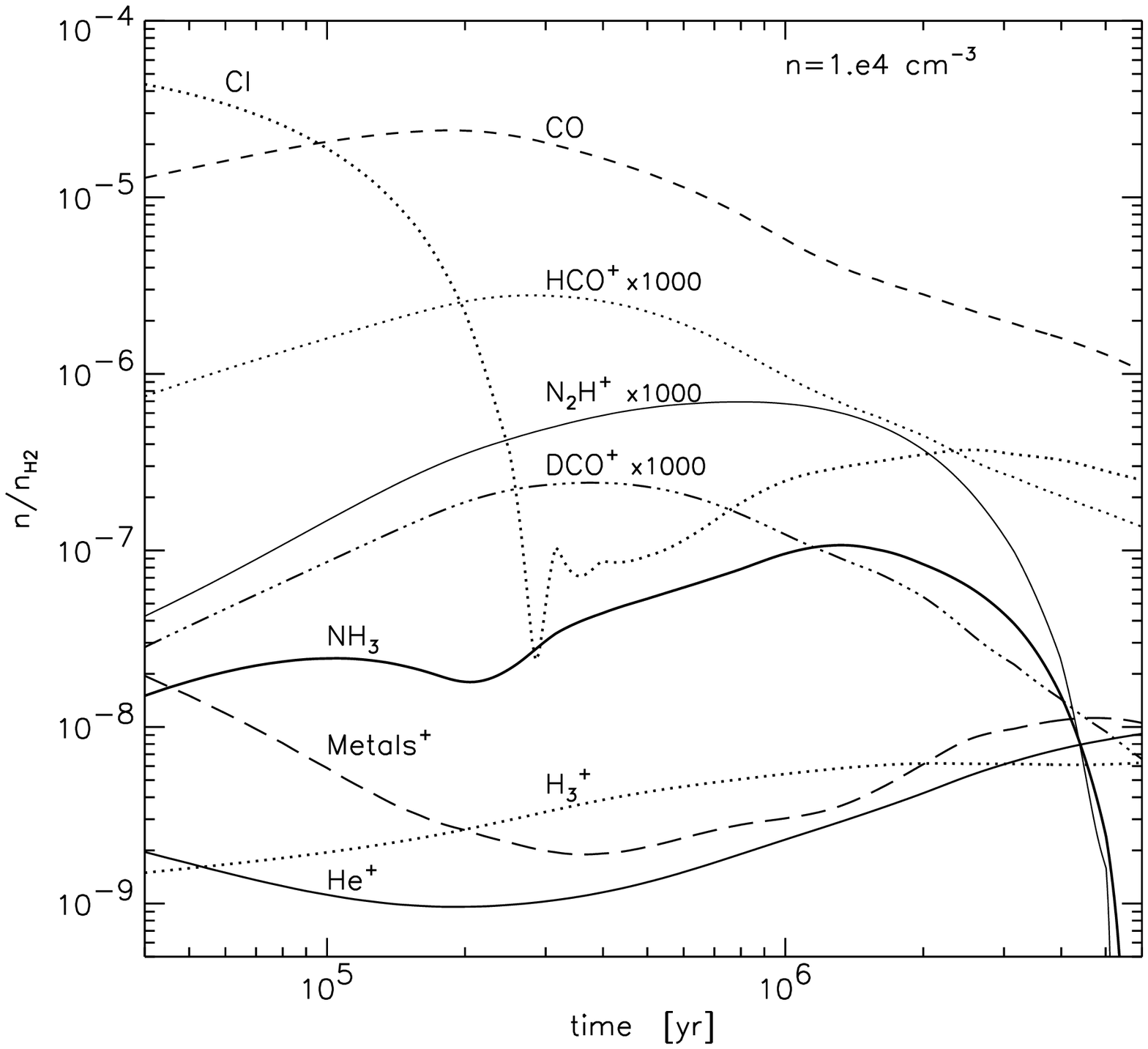}}
\caption[]{}
\label{fig4}
\end{figure}

{\bf Figure \ref{fig4}:} Time evolution of molecular abundances in the model
with $n= 10^4$~cm$^{-3}$ and \aav =5.5~mag. The abundances of HCO$^+$, 
N$_2$H$^+$ and DCO$^+$ have been multiplied by a factor of 1000. The dashed 
line labeled 'Metals$^+$' gives the sum of the abundances of C$^+$, Si$^+$,
Fe$^+$, N$^+$ and O$^+$.\\

\begin{figure}
\centerline{
\epsfxsize=9cm \epsfbox{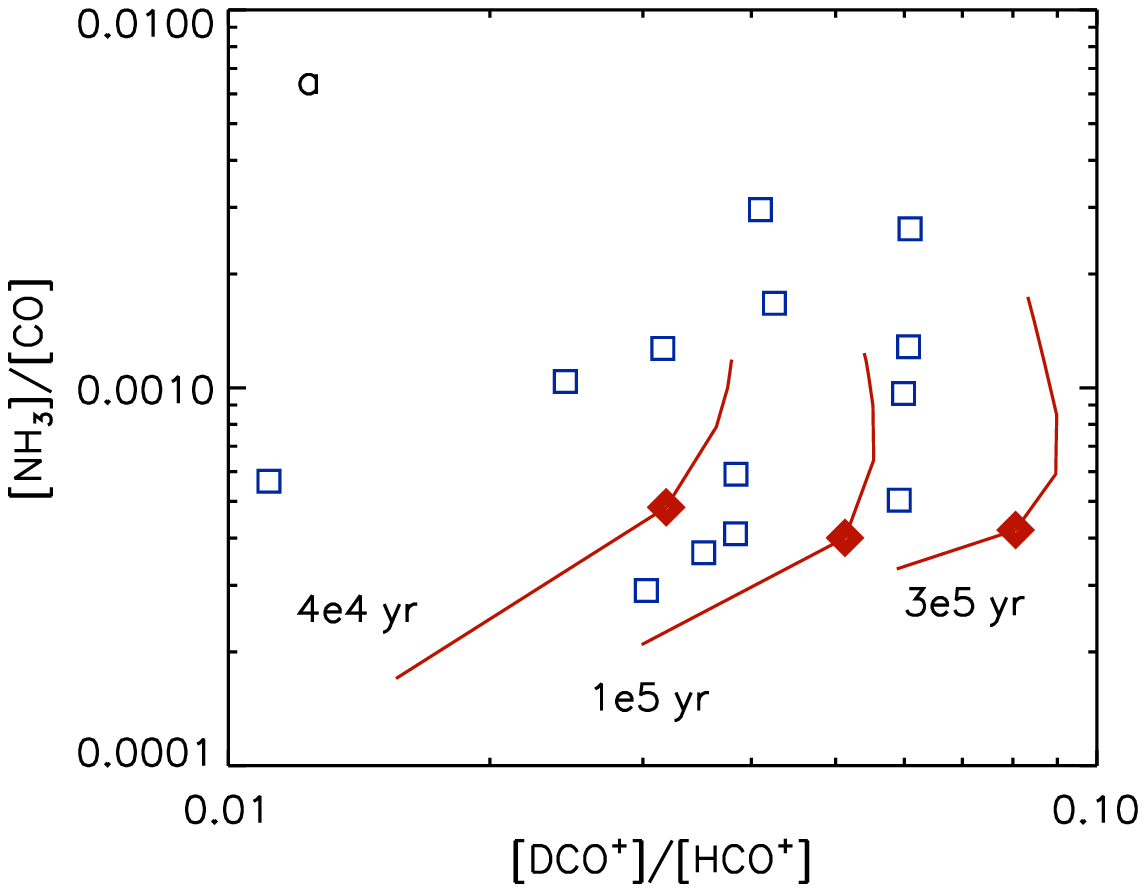}
\epsfxsize=9cm \epsfbox{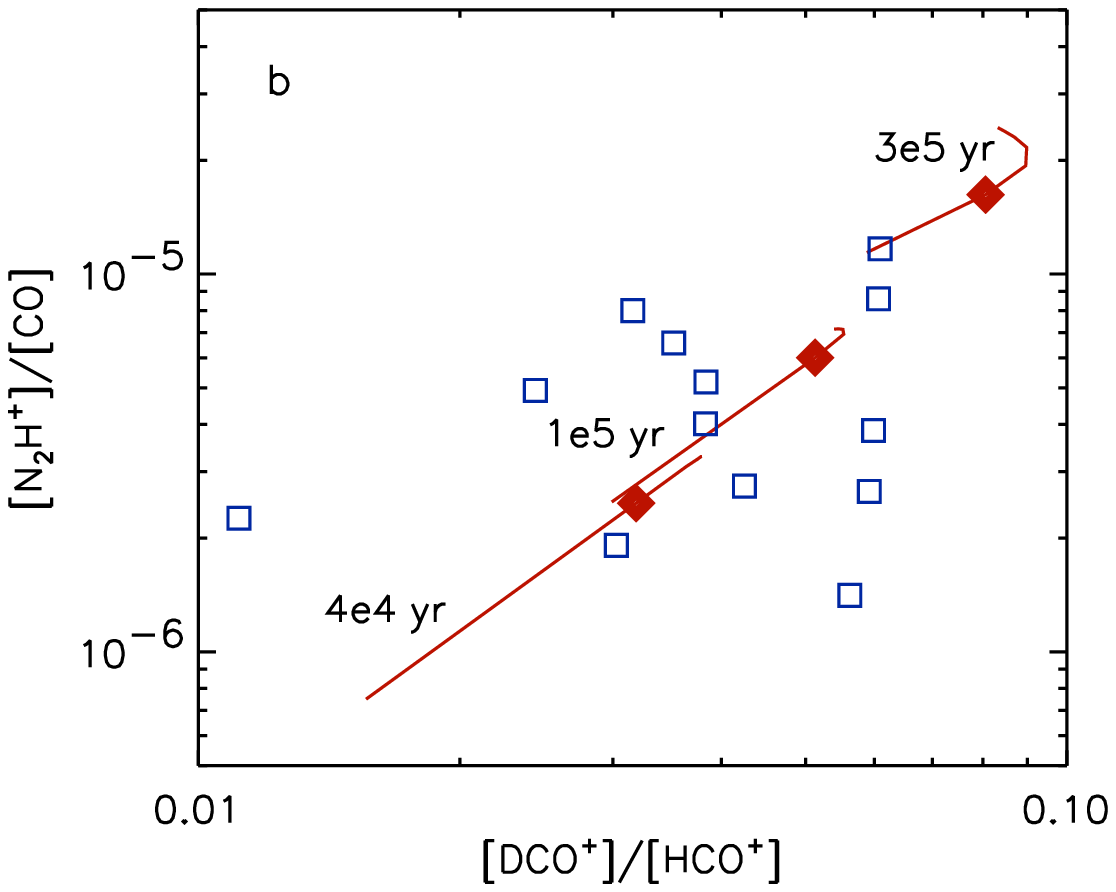}
}
\centerline{
\epsfxsize=9cm \epsfbox{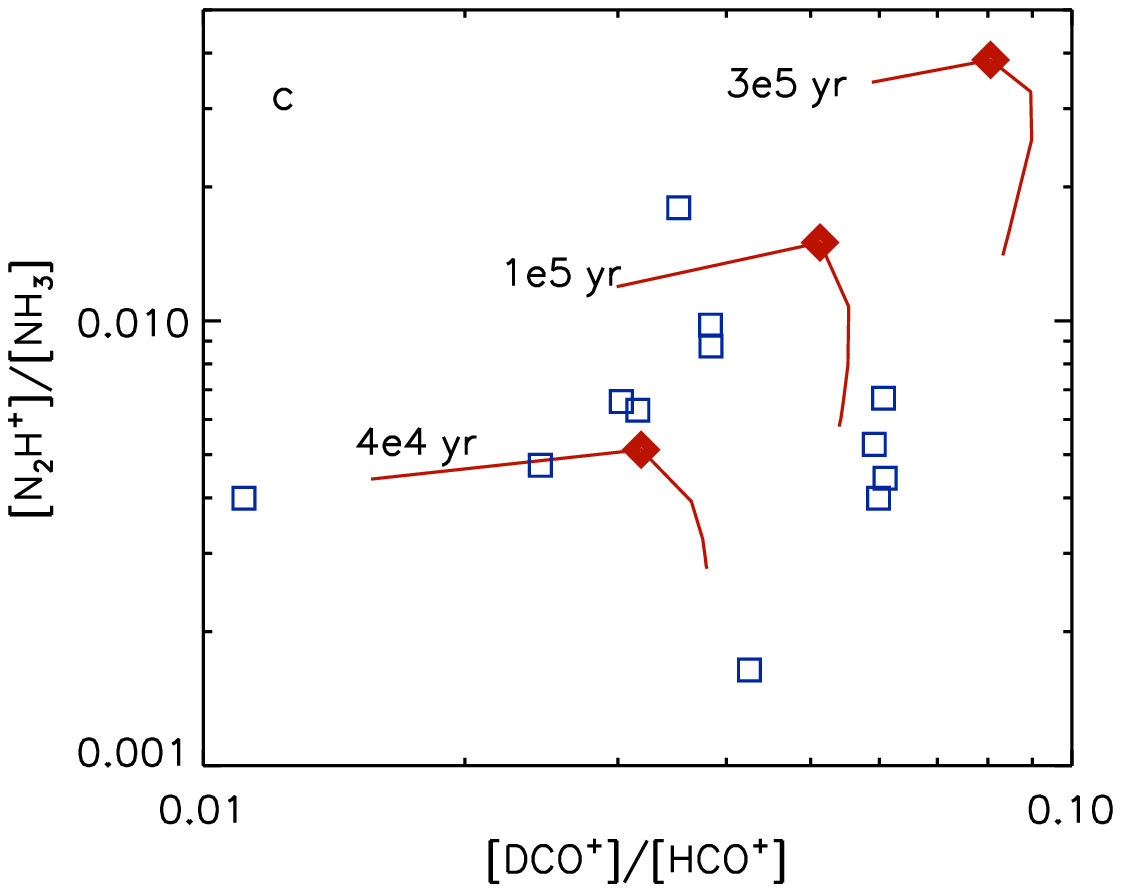}
\epsfxsize=9cm \epsfbox{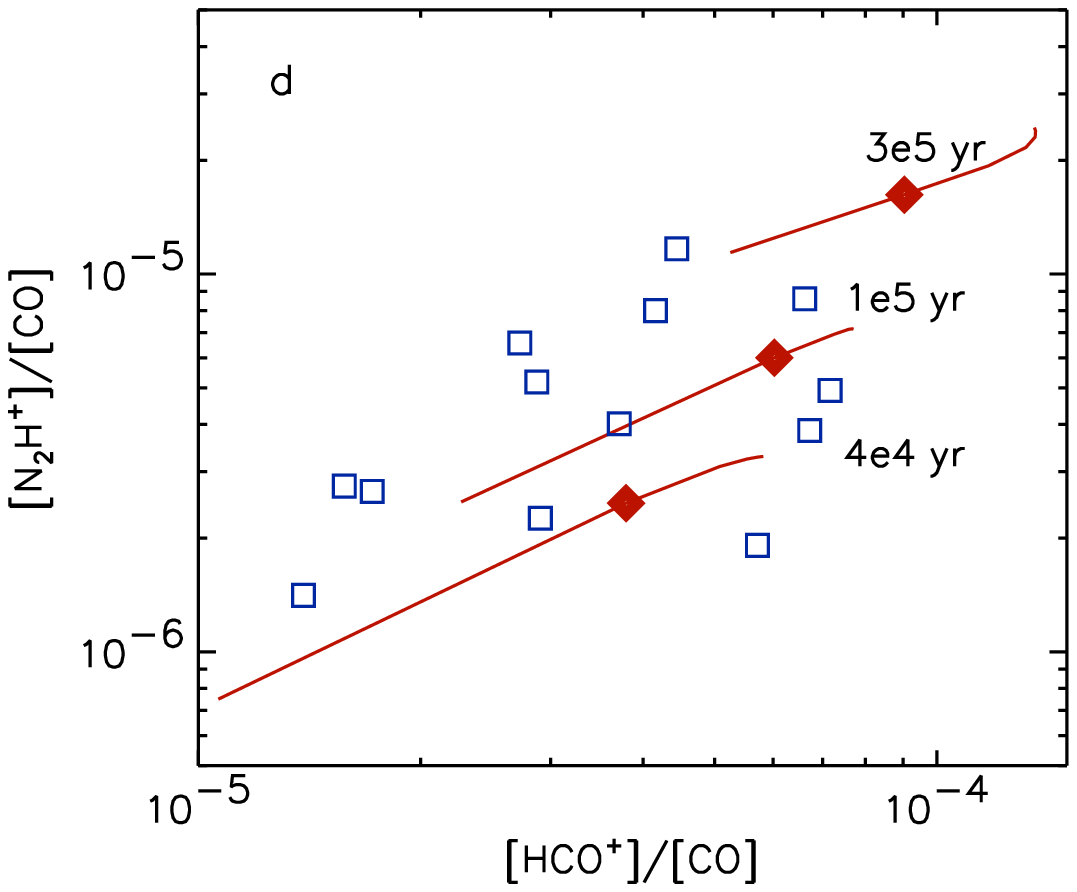}
}
\caption[]{}
\label{fig10}
\end{figure}

{\bf Figure \ref{fig10}:} Isochrones of models with $n= 10^4$~cm$^{-3}$.
Each curve represents the variation in abundances for different values of \aav
(ranging from \aav = 3 mag at the bottom of the curve to \aav = 6 mag at the top)
at a fixed time $4\times 10^4$~yr, $10^5$~yr and $3\times 10^5$~yr (solid lines). 
Squares correspond to a subsample of the starless cores from Butner, Lada and 
Loren (1995) and \cite{Williams+98} for which NH$_3$ and N$_2$H$^+$ 
abundances were given by Jijina, Myers and Adams (1999) and by 
\cite{Caselli+2002cores}. Filled diamonds correspond to \aav = 3.5 mag.
a) [NH$_3$]/[CO] versus \ratiox ; b) [N$_2$H$^+$]/[CO] versus 
\ratiox ; c) [N$_2$H$^+$]/[NH$_3$] versus \ratiox ; d) [N$_2$H$^+$]/[CO] 
versus [HCO$^+$]/[CO]. \\

\begin{figure}
\centerline{\epsfxsize=20cm \epsfbox{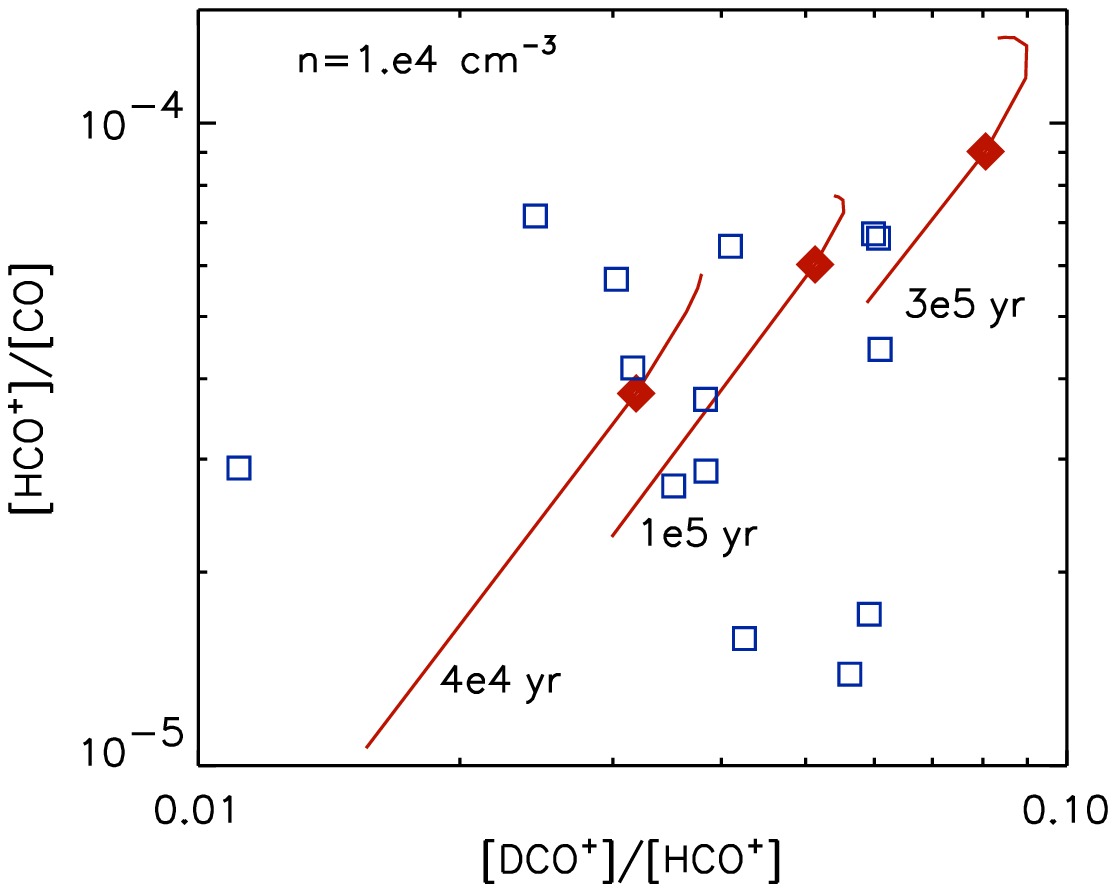}}
\caption[]{}
\label{fig5}
\end{figure}

{\bf Figure \ref{fig5}:} Isochrones of models with $n= 10^4$~cm$^{-3}$.
Each curve represents the variation in abundances for different values of \aav
(ranging from \aav = 3 mag at the bottom of the curve to \aav = 6 mag at the top)
at a fixed time $4\times 10^4$~yr, $10^5$~yr and $3\times 10^5$~yr (solid lines). 
Squares correspond to the starless cores from the samples in
Butner, Lada and Loren (1995) and \cite{Williams+98}. Filled diamonds 
correspond to \aav = 3.5 mag. \\

\begin{figure}
\centerline{\epsfxsize=20cm \epsfbox{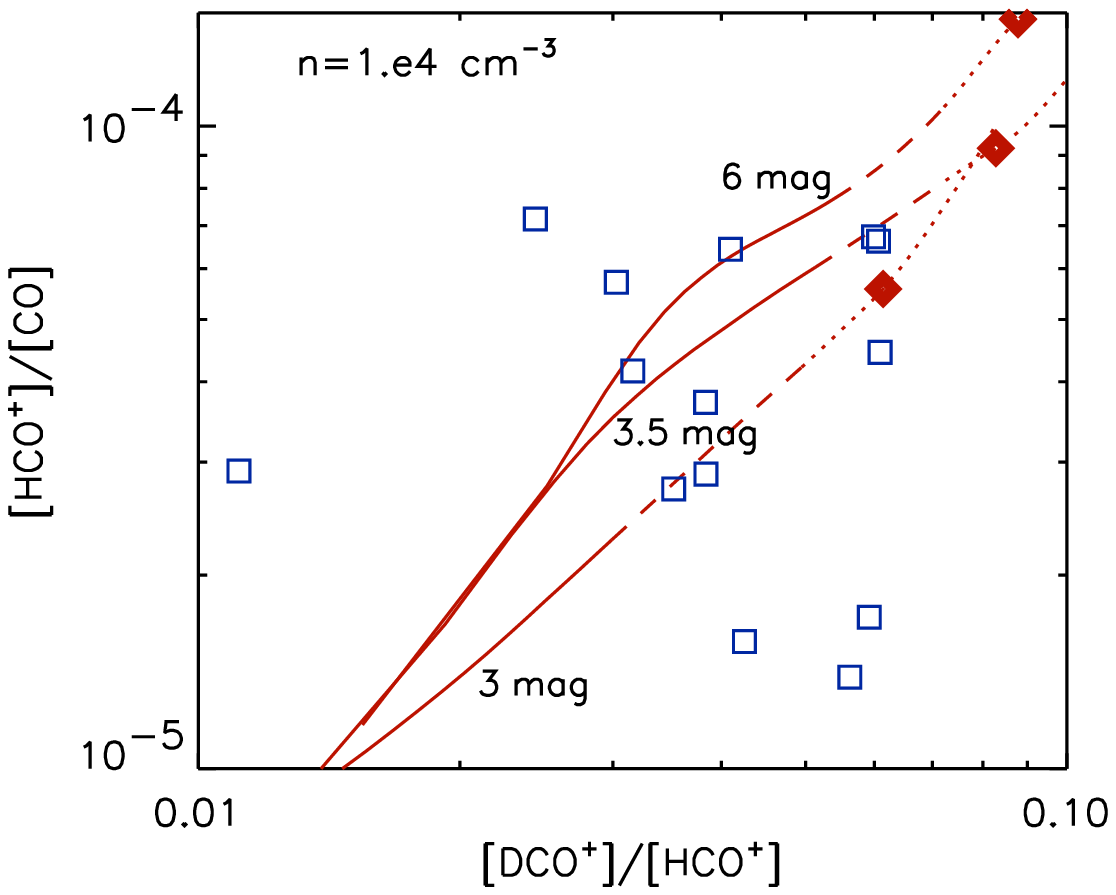}}
\caption[]{}
\label{fig6}
\end{figure}

{\bf Figure \ref{fig6}:} Time evolution tracks of models with 
$n= 10^4$~cm$^{-3}$, for \aav=3~mag, \aav=3.5~mag and \aav=6~mag
on the \rh\ plane. Each track is shown as a solid line for
an age up to $10^5$~yr, as a dashed line between $10^5$~yr and 
$2\times 10^5$~yr, and as a dotted line between $2\times 10^5$~yr 
and $10^6$~yr. Filled circles correspond to the free--fall time, 
$t=t_{\rm ff}=3.5\times 10^5$~yr. Squares correspond to 
the starless cores from the samples in Butner, Lada and Loren (1995) 
and \cite{Williams+98}.\\

\begin{figure}
\centerline{\epsfxsize=13cm \epsfbox{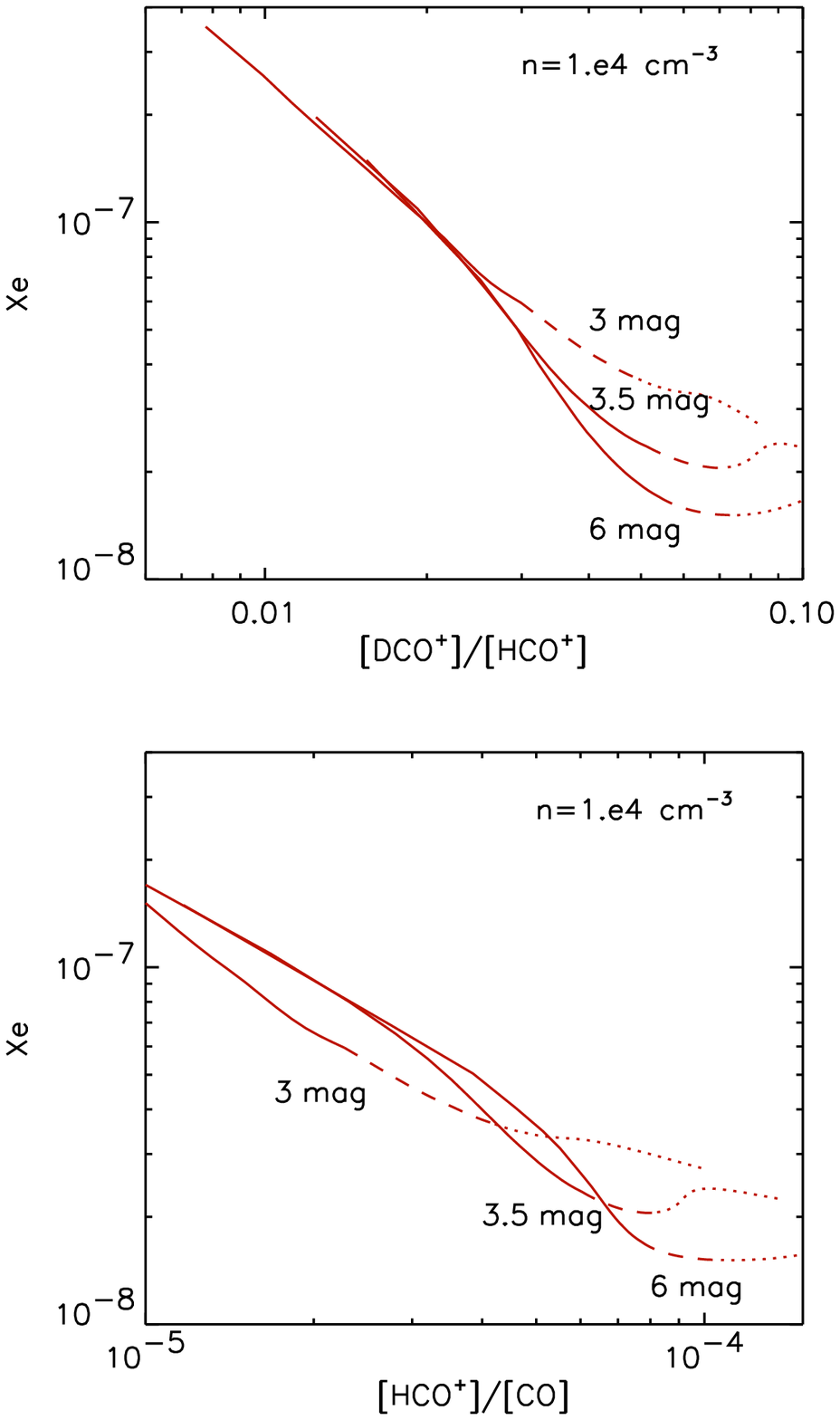}}
\caption[]{}
\label{fig7}
\end{figure}

{\bf Figure \ref{fig7}:} Tracks showing the time evolution 
of the electron abundance, \xe
for models with $n = 10^4$ cm$^-3$ and \aav=3, 3.5 and 6 mag
on the \rh\ plane.  Each track is shown as a solid line for
an age up to 10$^5$~yr, as a dashed line between 10$^5$ and
2 $times$ 10$^5$~yr, and as a dotted line between 
2 $\times$ 10$^5$~yr and 10$^6$~yr. \\

\begin{figure}
\centerline{\epsfxsize=20cm \epsfbox{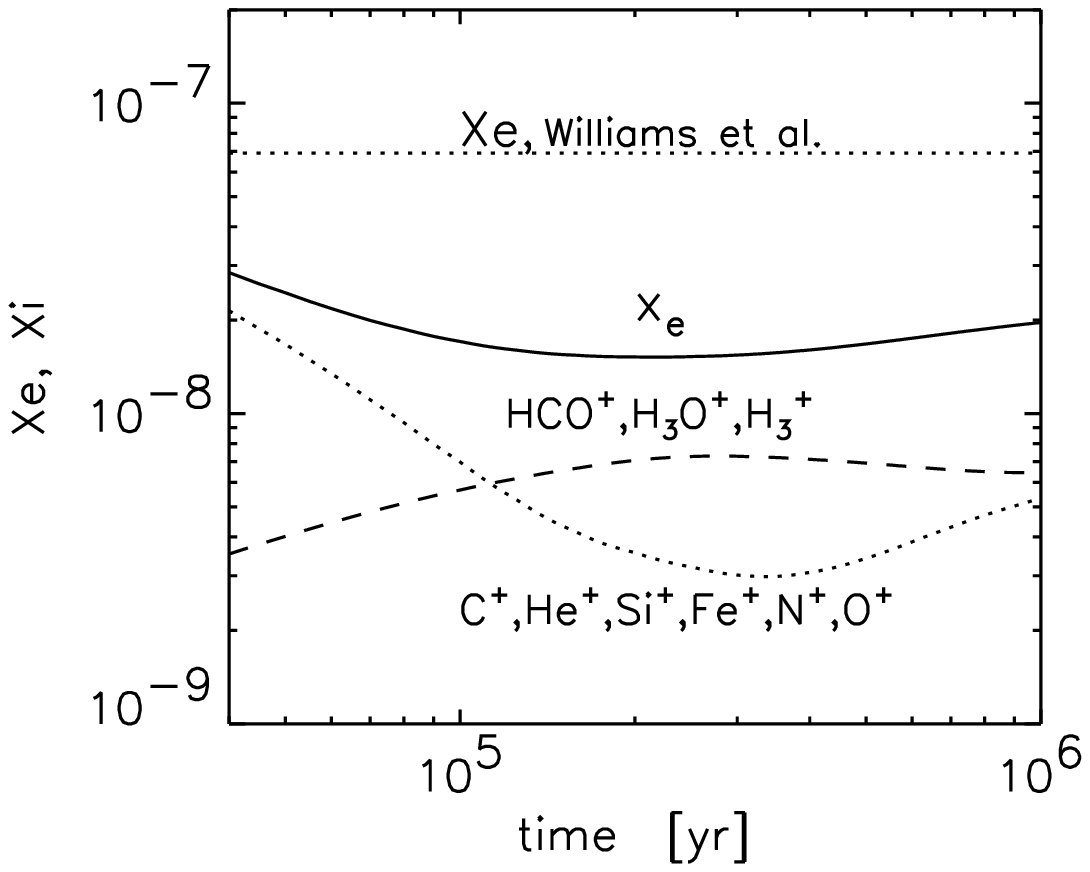}}
\caption[]{}
\label{fig8}
\end{figure}

{\bf Figure \ref{fig8}:} Ion and electron abundances versus time
in the model with $n= 10^4$~cm$^{-3}$ and \aav=5.5~mag.  The electron
abundance according to the results of \cite{Williams+98}
(\xe = $C_i n^{-1/2} \zeta_{\hbox{H}_2}^{1/2}$ where $C_i$ = 
2.0 $\times$ 10$^3$ cm$^{-3/2}$s$^{1/2}$) is shown for comparison.\\

\clearpage

\begin{table}
\begin{tabular}{l|c}
\hline
\hline
X & $n({\rm X})/n({\rm H}_{\rm tot})$ \\
\hline
H$_2$   &     0.495 \\                   
HD      &     1.6e-5  \\
N       &     2.14e-5  \\
O       &     1.76e-4  \\
Si      &     2.0e-8  \\
C$^+$   &     7.3e-5  \\
Fe$^+$  &     1.0e-8  \\
CO      &     0   \\
HCO$^+$ &     0  \\
DCO$^+$ &     0  \\
\hline
\end{tabular}
\caption{Initial chemical abundances relative to $n(\rm H_{tot})$,
where $n(\rm H_{tot})=2n({\rm H}_2)+n({\rm H})$.}
\label{t1}
\end{table}

\clearpage

\clearpage

\clearpage

\clearpage

\clearpage

\clearpage

\clearpage

\clearpage

\clearpage

\end{document}